\documentclass[%
 reprint,
superscriptaddress,
 amsmath,amssymb,
 aps,
]{revtex4-1}

\usepackage{graphicx}
\usepackage{dcolumn}
\usepackage{bm}

\usepackage{amsmath}
\usepackage{xfrac}

\newcommand{\td}{\mathrm{d}}
\newcommand{\Wmdm}{\Omega_{\mathrm{mdm}}}
\newcommand{\Wedm}{\Omega_{\mathrm{edm}}}
\newcommand{\vWedm}{\vec{\Omega}_{\mathrm{edm}}}
\newcommand{\vWmdm}{\vec{\Omega}_{\mathrm{mdm}}}

\begin{document}

\preprint{APS/123-QED}

\title{Frequency domain method of the search for the deuteron electric dipole moment in a storage ring with imperfections \\}

\author{Yury Senichev}
\email{y.senichev@fz-juelich.de}
\affiliation{Institut f\"ur Kernphysik, Forschungszentrum J\"ulich, 52425 Jülich, Germany}
\author{Alexander Aksentev}%
\affiliation{Institut f\"ur Kernphysik, Forschungszentrum J\"ulich, 52425 Jülich, Germany}%
\affiliation{National Research Nuclear University ``MEPhI,'' 115409 Moscow, Russia}%
\author{Andrey Ivanov}%
\affiliation{Faculty of Applied Mathematics and Control Processes,
	St. Petersburg State University, 198504 Petersburg, Russia}%
\author{Eremey Valetov}%
\affiliation{Department of Physics and Astronomy, Michigan State University, East Lansing, Michigan 48824, USA}%

\date{\today}

\begin{abstract}
The method is based on four fundamental features. The total spin precession frequency in the vertical plane due to the electric and magnetic dipole moments in an imperfect ring in a vertical plane is measured. The position of the ring elements is unchanged from clockwise to counter-clockwise operation. The calibration of the effective Lorentz factor using the polarization precession frequency measurement in the horizontal plane is carried out alternately in each CW and CCW operation. And the approximate relationship between the frequencies of the polarization precession in different planes is set to exclude them from mixing to the vertical frequency of the expected EDM signal at a sensitivity level approaching  $10^{-29}$ $e\cdot cm$. 
\begin{description}
\item[PACS numbers]
13.40.Em, 11.30.Er, 29.20.Dh, 29.27.Hj
\end{description}
\end{abstract}

\maketitle


One of the essential problems of modern physics is the baryon asymmetry of the universe, which indicates the prevalence of matter over antimatter~\cite{Canetti}.In addition, the cosmic detectors PAMELA and AMS, whose purpose is to search for antimatter, have yet to find a significant amount of it in the universe ~\cite{Aguilar}. A new idea claiming that one of the reasons for the baryon asymmetry is the breaking of CP invariance emerged soon after its discovery. A. Sakharov established the necessary conditions for baryogenesis (the initial creation of baryons) in 1967 ~\cite{Sakharov}. Many theories beyond the Standard Model (SM) have been proposed -- all of them new physics theories -- that are able to remove the difficulties encountered in the SM but have yet to be proven in experiments. One of the possible signatures for the breaking of CP invariance is the existence of non-vanishing electric dipole moments (EDM) of elementary particles.

The idea of searching for the electric dipole moment (EDM) of the proton and the deuteron using polarized beams in a storage ring is based on the frozen spin (FS) method and was originally proposed at Brookhaven National Laboratory (BNL)~\cite{Farley}. The concept of the “frozen spin” lattice consists of deflectors with electric and magnetic fields incorporated in one element so that the spin of the reference particle is always orientated along the momentum. Under these conditions, the signal growth of the assumed electric dipole moment, a rotation into the vertical plane, is maximized. This is clearly evident from the Thomas–Bargmann–Michel–Telegdi equation:
\begin{align}
	\frac{\td\vec{S}}{\td t} & = \vec{S}\times \left(\vWmdm + \vWedm\right), \notag\\
	\vWmdm &= \frac{e}{m\gamma}\left \{\left(\gamma G + 1\right)\vec{B} - \left(\gamma G + \frac{\gamma}{\gamma+1}\frac{\vec{\beta}\times\vec{B}}{c}\right)\right \} , \notag\\
	\vWedm &= \frac{e\eta}{2m}\left(\vec{\beta}\times\vec{B} + \frac{\vec{E}}{c}\right);~ G = \frac{g-2}{2}\label{eq:Wedm}
\end{align}
where $G$ is the anomalous magnetic moment, $g$ is the gyromagnetic ratio, $\Wmdm$ is the spin precession frequency due to the magnetic dipole moment (hereinafter referred to as MDM precession), $\Wedm$ is the spin precession frequency due to the electrical dipole moment (hereinafter referred to as EDM precession), and $\eta$ is the dimensionless coefficient defined by the relation $d = \eta e \hbar/4mc$. 

It is advantageous to implement the frozen spin method in a purely electrostatic storage ring ($\vec{B} = 0$) with electrostatic deflectors confining the beam to a closed orbit. The advantages of purely electrostatic storage ring are especially evident at the ``magic'' energy, when
\begin{equation}\label{eq:MagicCond}
	G - \frac{1}{\gamma^2_{mag}-1} = 0,
\end{equation}
and the polarization, initially oriented in the longitudinal direction, rotates in the horizontal plane with the same frequency as the momentum $\Omega_p$ , i.e. $\Wmdm - \Omega_p= 0$ . 

However, this method cannot be used for deuterons with the negative anomalous magnetic moment $G = -0.143$, which follows from condition~\eqref{eq:MagicCond}. Therefore, the only possible method in this case is a storage ring with both electric and magnetic fields~\cite{AGS_proposal}. It was proposed that a longitudinally polarized deuteron beam of 1 GeV/c total momentum could be stored in an electromagnetic storage ring where the MDM precession is minimized with respect to momentum $\Wmdm^p$ . This can be done by applying a radial electric field $E_r$  of a magnitude to balance the vertical magnetic field  $B_v$ contribution to $\Wmdm^p$ relative to $\Omega_p$, as shown in Eq.~\eqref{eq:Wedm}, which requires: 
\begin{equation}\label{eq:ErBv}
E_r= \frac{GBc\beta\gamma^2}{1-G\beta^2\gamma^2}\approx G B_v c\beta\gamma^2
\end{equation}	 

Thus, in both proton and deuteron rings it is possible for the concept of a frozen spin to be realized. In this sense, there is a general concept of how to construct such a ring, but this is realized with the help of different types of deflectors. These differences do not play an essential role with respect to the spin-orbital motion, so long as the motion of the reference particle in the perfect ring without misalignments is considered.  

The presence of errors in the installation of the elements (imperfections) of the ring leads to the appearance of vertical $E_v$  and radial $B_r$   components of the electric and magnetic fields, respectively. They both change the spin components in the vertical plane where the EDM signal is expected, and create the systematic errors that produce a ``fake EDM'' signal

To solve this problem in the case of a proton beam, it was suggested that the procedure of simultaneously injecting two beams in the ring into the two opposite directions, clockwise (CW) and counterclockwise (CCW)~\cite{AGS_proposal}, be used. . If in CW direction the deviation of the spin in the vertical plane due to MDM and EDM is added, then it is subtracted in CCW direction. By subtracting the CW and CCW results, the EDM can be separated from a systematic error arising due to MDM. However, this method is only possible in a purely electrostatic ring, since the deflecting force of the electric field does not depend on the direction of the charged particle and both beams can be simultaneously present in the ring. In the case of a deuteron ring, the magnetic component of the Lorentz force depends on the direction of motion, which therefore means that the polarity of the magnetic field needs to be changed when the direction of injection is different. 

There is therefore a global problem regarding how to restore the magnetic field and trajectory of the beams after reversing the polarity, since the field recovery error should not exceed the expected EDM value. The difference in the orbit lengths for CW and CCW beams leads to an unmeasured difference in MDM frequencies. This problem effectively ends the possibility of investigating deuteron EDM in the ring. This is the main argument behind the notion that a different approach is required for the deuteron ring. The inability to identify the orbit length of the CW and CCW beams gives priority to  the method described below in the case of the search for the EDM of protons as well.  

In addition, the erroneous radial magnetic $B_r$   and vertical electric $E_v$   components act differently on the spin due to the imperfections of the ring elements. Even if we assume that the vertical component of the Lorentz force averaged over the ring $\overline{F_v}$   is equal to exactly zero due to the ideal adjustment of fields in elements of the ring to provide stable motion~\cite{AGS_proposal},
\begin{equation}\label{eq:Fav}
\overline{F_v}= e\left(\overline{vB_v}-\overline{E_r}\right)
\end{equation} 
we would still observe a non-zero rotation of the spin in the vertical plane, that is to say the “fake EDM” signal. Let us assume that $n$  number of arbitrary elements of the size $L$  are installed on the ring with the rms vertical error $\left\langle{\delta h} \right\rangle$ , and that the conditions in  (4) are fulfilled. From (1) and (4) we can obtain the standard deviation of MDM spin precession frequency in a vertical plane around the radial axis: 
\begin{equation}\label{eq:Omega av}
\left\langle{\Omega_{r,\mathrm{edm}}} \right\rangle= 
\frac{e}{m\gamma}\frac { G + 1}{\gamma}\frac{\left\langle{B_r}\right\rangle}{\sqrt{n}} 
\end{equation}
where $\left\langle{B_r}\right\rangle $   is the rms value of the radial magnetic field. Here, we no longer subtract the momentum frequency from $\Omega_{r,\mathrm{edm}}$   as was the case in the horizontal plane, since there is no momentum rotation in a vertical plane and the frozen condition (2) is irrelevant. It is worth noting that such a high sensitivity to errors occurs in a purely electrostatic ring as well. The value of the radial component of the field  $\left\langle{B_r}\right\rangle=B_v\left\langle{\delta h} \right\rangle/L$  is thus determined by the slope of the magnet $\left\langle{\delta h} \right\rangle/L$   in a transverse plane around the longitudinal axis and the vertical component of the magnetic field $B_v$. If we assume a more or less realistic value of rms error installation of an arbitrary magnet $\left\langle{\delta h} \right\rangle =100\mu m$, the polarization  precession frequency in the vertical plane will be of the order of $\left\langle{\Omega_{r,\mathrm{edm}}}\right\rangle\approx   50$ rad/sec when the size of the magnets is $L\approx1$ m and the total number of elements on the ring is $n\approx100$. The numerical simulation of spin-orbital motion in the design lattice using the COSY INFINITY code~\cite{Berz} gives a frequency value about 1.5 times lower, which indicates that the formula (5) can be applied to estimate the order with sufficient success. If we compare the frequency of polarization precession due to an assumed EDM with the frequency caused by MDM in a vertical plane, the latter, which is mixed with the EDM, initiates a fake signal, thus resulting in the systematic errors that limit the accuracy of determining the EDM in our future experiment. We can determine the polarization precession frequency in the vertical plane caused by the assumed EDM to be  $\Omega_{\mathrm{edm}}=10^{-8}$ rad/sec by substituting the ring design parameters $B_v\approx 4$ T and $E_r\approx$ 12 MV/m in (1)~\cite{ICAP2015}. It is important to note here that in the BNL experiment proposal for the EDM search~\cite{AGS_proposal}, the aim is to measure the value of the vertical component of the spin initiated by EDM at the level of $S_v\approx 10^{-6}$  , which is reached in 1,000 seconds. Evidently, with this approach, we must lower the MDM contribution to a level below the expected value of the EDM. In accordance with this requirement, the accuracy of the geodetic installation must be increased by more than ten orders of magnitude down to an unrealistic value of $10^{-14}$  m. We are therefore unable to measure the accumulated EDM signal by observing the growth of the vertical component of spin as suggested in~\cite{Farley,AGS_proposal}, since the spin rotates in the plane where we expect to see EDM at an incomparably higher speed due to MDM.

Therefore, we propose measuring not the magnitude of the vertical component of spin but rather the total precession frequency of the spin due to the EDM and MDM. In order to separate the EDM signal from the sum signal, an additional condition is required, namely the total spin frequency in the experiment is measured with a clockwise direction of the beam 
\begin{equation}\label{eq:Omega CW}
\Omega_{\mathrm{CW}}=\Omega_{r,\mathrm{mdm}}^{\mathrm{CW}}+\Omega_{\mathrm{edm}}
\end{equation}  
and compared with  counter-clockwise measurements 
\begin{equation}\label{eq:Omega CCW}
\Omega_{\mathrm{CCW}}=-\Omega_{r,\mathrm{mdm}}^{\mathrm{CCW}}+\Omega_{\mathrm{edm}}
\end{equation}
The sum of the frequencies of these two signals 
\begin{equation}\label{eq:Omega CW_CCW}
\Omega_{\mathrm{edm}}=0.5\left(\Omega_{\mathrm{CW}}+\Omega_{\mathrm{CCW}}\right)+0.5\left(\Omega_{r,\mathrm{mdm}}^{\mathrm{CCW}}-\Omega_{r,\mathrm{mdm}}^{\mathrm{CW}}\right)
\end{equation}
allows us to identify the frequency of the EDM signal, which in turn provides us with the EDM value. However, there are four problems here, all of which require special consideration.
 
First, we see in (8) that the accuracy of the frequency measurements of $\Omega_{\mathrm{CW}}$ and $\Omega_{\mathrm{CCW}}$   determines the precision of the EDM measurement. In~\cite{Eversmann}, it is shown that the relative accuracy of the polarization precession frequency measurement, $10^{-10}$ to $10^{-11}$,   is achievable even when the polarimeter measurement frequency is much less than the polarization precession frequency. In our case, we have an inverse relationship between the polarimeter rate and the measured spin frequency, which extends the range of frequencies where statistical estimates are legitimate. For an absolute statistical error of measuring a frequency of the spin oscillation, we can use $\sigma_{\Omega}=\delta\epsilon_{A}\sqrt{24/N}/T$, where $N$  is the total number of recorded events, $\delta\epsilon_{A}$   is the relative error in measuring the asymmetry, and $T\approx 1000$  sec is the measurement duration~\cite{Aksentyev}. If we assume a beam of $10^{11}$  particles per fill and a polarimeter efficiency of one percent, this leads to an absolute error of frequency measurement of  $\sigma_{\Omega}=2\cdot10^{-7}$ rad/sec. With a nominal accelerator beam time of 6,000 hours per year, we can reach $\sigma_{\Omega}=2\cdot10^{-9}$ rad/sec during one year. If we take into account that formula (1) with the EDM $d_d\approx 10^{-30} e\cdot cm$  gives a value of the spin precession frequency of $\Omega_{\mathrm{edm}}\approx 10^{-8}$  , we can state that the accuracy for the frequency of $\sigma_{\Omega}=1.4\cdot10^{-9}$   is satisfactory and sufficient for reaching a sensitivity of $d_d\approx 10^{-30}e\cdot cm$   ( where $\eta\approx2\cdot10^{-15}$ ).

Second, the main idea behind using CW and CCW procedures is that the contribution of the MDM spin rotation is the same for both CW and CCW directions. In an ideal scenario, the difference $\Omega_{r,\mathrm{mdm}}^{\mathrm{CCW}}-\Omega_{r,\mathrm{mdm}}^{\mathrm{CW}}$   is zero. However, this is not exactly the case. In reality, we do not know how accurately the field is recovered after a change of polarity, that is to say whether the energy of the beam is the same or not. Furthermore, the CW and CCW beam trajectories may  have different orbit lengths, which in turn contribute to the MDM spin precession frequency. We must therefore reformulate the global problem regarding how to restore the conditions for the equal contribution of the two MDM spin rotations after a change in the polarity of magnetic field (no change for electrical field) in the plane where we will measure the EDM. We expect to achieve a difference $\Omega_{r,\mathrm{mdm}}^{\mathrm{CCW}}-\Omega_{r,\mathrm{mdm}}^{\mathrm{CW}}$ that is smaller than the expected EDM precession frequency   $\Omega_{\mathrm{edm}}$ (see eq. 8). In this regard, we will undertake two procedures. The first follows from the study of the suppression of the spin decoherence~\cite{IPAC2013}, where we reached a very important conclusion, namely that two particles that  have different orbits and different initial conditions in their transverse planes as well as different initial energy deviations are assumed to be the same – from the point of view of the spin behavior – if they have the same spin tune. The latter is reached when the lengths of the particle orbits are equal. We refer to this as the conditions of zero decoherence of the spin precession. Equation (8) quoted from~\cite{IPAC2013} shows this dependence:
\begin{equation}\label{eq:delta p}
\Delta\delta_{eq}=\frac{\gamma_{s}^{2}}{\gamma_{s}^{2}\alpha_0-1}\left[\frac{\delta_{m}^{2}}{2}\left(\alpha_{1}-\frac{\alpha_{0}}{\gamma_{s}}^{2}+\frac{1}{\gamma_{s}^{4}}\right)+\left(\frac{\Delta L}{L}\right)_{\beta}\right]	
	\end{equation}
	
where $\Delta\delta_{eq}$  is the deviation of the equilibrium level (average value) of the momentum due to the orbit increasing in length in the transverse plane $\left(\frac{\Delta L}{L}\right)_{\beta}$   for a synchrotron oscillation with amplitude $\delta_{m}$ of momentum. Values $\alpha_{0}$ and $\alpha_{1}$  are the zero and first order momentum compaction factors, while $\gamma_{s}$  is the Lorentz factor of the synchronous particle. 

The new equilibrium energy,
\begin{equation}\label{eq:effec gamma}
\gamma_{eff}=\gamma_{s}+\beta_s^{2} \gamma_s\cdot\Delta\delta_{eq}	
\end{equation}
will hereinafter be referred to as the effective Lorentz factor. This parameter is named as such because $\gamma_{eff}$ includes three spatial coordinates and completely determines the frequency of spin precession in all three planes. The orbit length of each particle is adjusted by the sextupoles, leading to a dependence of the action of the sextupole field on the amplitude of transverse oscillations and the energy deviation from the reference particle. We can now apply this important conclusion to the beams moving in opposite CW and CCW directions, namely that the beams are identical in terms of the spin behavior if they have the same effective Lorentz factor averaged over all particles in beam. This means that the problem of finding the multiparameter dependence of spin precession on fields and 3D trajectories is reduced to the search for a dependence on the effective gamma. This ensures it is no longer necessary to obtain a coincidence of trajectories, but instead only requires the condition of equality $\gamma_{eff}$   for the CW and CCW beams. This approach saves the whole idea of searching for an EDM in a storage ring.

Third, if we assume that there are two rings with a direct (CW) and reverse sequence of elements (CCW) with a changed polarity of the magnetic field, the similarity of these rings under the beam stability condition (4) is only that the position of all elements on the ring and, consequently, the relation between the values of the vertical and radial components of the field remains unchanged
\begin{equation}\label{eq:element position}
B_r/B_v=const\ \text {and}\ E_v/E_r	
\end{equation}
Here, we should take into account the two facts mentioned above: first, we will not change the polarity of the electric field, leaving it unchanged during the transition from CW to CCW, and second, when the energy rises, the condition of equilibrium for the particles will be maintained by changing the magnetic field only. In our case this is a change in the magnetic field from 0.3 to 0.45 Tesla. Therefore, we will in the future only focus our attention on the magnetic field. Irrespective of this circumstance, it is unlikely that the reverse trajectory will coincide with the direct trajectory, which can be a reason for having a different orbit length and, hence, different Lorentz factor values   that determine the spin precession frequency in all planes. Therefore, before changing the polarity, we must calibrate the gamma $\gamma_{eff}$   close to the value $\gamma\approx\gamma_s$   using the precession frequency measurements of the spin in the horizontal plane where there is no EDM signal before restoring the same $\gamma_{eff}$   in the ring with the reverse sequence of elements. For such a calibration, we need to reduce the spin oscillation in the vertical plane to a low value by introducing a spin rotator 1 m long with $\vec{E}\times \vec{B}$ transverse magnetic  and  electric fields  in the order of 0.1 mT and 100 V/cm respectively.  The spin rotator is installed in a straight section and provides zero Lorentz force on axis. The value of this field does not affect the calibration of the effective Lorentz factor. Here, we are aiming to observe how to slow down the spin rotation in the vertical plane which does not mean that we are introducing the magnitude of the fields, as a parameter requiring precise determination. The sole purpose of introducing a rotator is to ensure that the relative contribution of the vertical frequency into the horizontal frequency is less than the calibration accuracy required, namely $ 10^{-9}$  . Since they add up as squares of frequencies, this can be easily achieved. The transverse spin rotator is switched on only for the time of calibration of $\gamma_{eff}$ the   in the CW ring and for the time of its recovery in the CCW ring. We are able to calibrate the frequency, $\gamma_{eff}$ , with the above-mentioned absolute value of errors for one beam fill of $\sigma_{\Omega}\approx10^{-7}$ rad/sec and $\sigma_{\Omega}\approx10^{-9}$ rad/sec with one year of running. Taking into account the constant relation between the vertical and the radial components of field (11), this means that in the case of CCW we have a ring identical to the CW ring in terms of spin behavior, and we can obtain a zero value of $\Omega_{r,\mathrm{mdm}}^{\mathrm{CCW}}-\Omega_{r,\mathrm{mdm}}^{\mathrm{CW}}$    with an accuracy of $\approx 10^{-9}$.

Finally, we will consider the fourth important aspect in the proposed procedure for measuring the EDM. This problem concerns the fact that the spin oscillation in any of the planes includes an uncontrollable spin oscillation in the other two planes. The solution of equation (1) under the initial conditions for horizontal, vertical ($S_x=0, S_y=0$)  , and longitudinal  $S_z=1$ components  can be formulated as shown here:
\begin{align}
S_x=\frac{\Omega_x\Omega_z}{\Omega^{2}}\left(1-\cos\Omega t\right)-\frac{\Omega_y}{\Omega}\sin\Omega t, \notag\\
S_y=\frac{\Omega_y\Omega_z}{\Omega^{2}}\left(1-\cos\Omega t\right)+\frac{\Omega_x}{\Omega}\sin\Omega t, \notag\\
S_z=\frac{\Omega_z^{2}}{\Omega^{2}}\left(1-\cos\Omega t\right)+\cos\Omega t,
\end{align}
where $\Omega_x=\Omega_{B_r}+\Omega_{edm}$   and $\Omega_z=\Omega_{B_z}$  arise due to MDM rotation in the imperfect ring and the EDM. $\Omega_{y}=\Omega_{B_{v},E_{r}}$  is the MDM spin rotation relative to the momentum in the leading magnetic and electric fields, and $\Omega=\sqrt{\Omega_x^{2}+\Omega_y^{2}+\Omega_z^{2}}$   is the modulus of the three-dimensional frequency. As mentioned above, we will measure the precession frequency of the spin in a vertical plane in order to study the behavior of the oscillating part of $\widetilde{S_y}$ , the solution to which is as follows:
\begin{align}
\widetilde{S_y}=\sqrt{\left(\frac{\Omega_y\Omega_z}{\Omega^{2}}\right)^{2}+\left(\frac{\Omega_x}{\Omega}\right)^{2}}\sin\left(\Omega t+\phi\right), \notag\\
\phi=\arctan\left(\frac{\Omega_y\Omega_z}{\Omega_x\Omega}\right),
\end{align}
Since the amplitude and the phase of the signal do not affect the measurement, we are only interested in the frequency:
\begin{equation}\label{eq:Omega}
\Omega=\sqrt{\left(\Omega_{edm}+\Omega_{B_r}\right)^{2}+\Omega_{B_v,E_r}^{2}+\Omega_{B_z}^{2}}	
\end{equation}
Assuming that, in accordance with the frozen spin concept, we maintain the spin along the momentum $\Omega_{B_v,E_r}<<\Omega_{B_r}$  and $\Omega_{B_z}<<\Omega_{B_r}$  , the latter expression is realized by installing a longitudinal solenoid one meter long on a straight section with a magnetic field of about $\approx 10^{-6}$  Tesla, which can be formulated as follows:
\begin{equation}\label{eq:Omega2}
\Omega=\left(\Omega_{edm}+\Omega_{B_r}\right)\cdot\left[1+\frac{\Omega_{B_v,E_r}^{2}+\Omega_{B_z}^{2}}{2\left(\Omega_{edm}+\Omega_{B_r}\right)^{2}}\right]	
\end{equation}
According to this equation, the restriction occurs at the values of $\Omega_{B_v,E_r}$   and $\Omega_{B_z}$ , which should have less of an effect on the total frequency $\Omega$   than the EDM:
\begin{equation}\label{eq:Omega3}
\frac{\Omega_{B_v,E_r}^{2}+\Omega_{B_z}^{2}}{2\Omega_{B_r}}<\Omega_{edm}\end{equation}
If we evaluate these requirements numerically, we see how difficult it is to implement them technically. For instance, if $\Omega_{B_r}\approx$ 50 rad/sec and $\Omega_{edm}\approx10^{-8}$  rad/sec, then $\Omega_{B_v,E_r}^{2}+\Omega_{B_z}^{2}<10^{-6}$   or both must be $\Omega_{B_v,E_r}, \Omega_{B_z}\sim10^{-3}$  . This means that at the spin coherence time $t_{SCT}\sim$ 1000 sec, the spin rotation should not exceed $\Omega_{B_v,E_r}\cdot t_{SCT}\sim$  1 rad and $\Omega_{B_z}\cdot t_{SCT}\sim$  ~1 rad, which is easily achievable both for $\Omega_{B_v,E_r}$   due to the calibration of energy and for $\Omega_{B_z}$   due to the introduction of a longitudinal solenoid. As in the case of a transverse spin rotator, the field in the longitudinal solenoid does not need to be known exactly, since it is only needed to satisfy equation (16), which is an approximation. We can therefore conclude that the imperfections of ring elements, which previously played a limiting role in the measurement of EDM, now provide a pure precession of the spin in the vertical plane, where we will measure the EDM.

In this article we have considered the contribution of systematic errors arising from the improper installation of deflectors and the requirement of changing the polarity of the field in them. However, the displacement of the magnetic quadrupoles can also lead to the appearance of a dipole field component that induces a fake signal, which requires an identity for CW and CCW. However, at least two solutions exist here. The first is to use electrostatic quadrupoles. The second is to use optics that does not require switching the polarity of the magnetic field in magnetic quadrupoles.

It should be noted that the idea of measuring the EDM by introducing a transverse coil and measuring the spin precession in the vertical plane was proposed in the wheel concept by I. Koop~\cite{Koop}. It differs from the method considered here. The wheel method uses a special horizontal coil. Assuming the field is calibrated in the coil, it measures the beam offset versus field values in the coil. Nevertheless, the problem with systematic errors from the presence of misalignments remains unsolved. In addition, the method requires special optics of the ring with a weak focusing in the vertical plane. This entails certain difficulties in the beam dynamics.

In this Letter, we described the frequency domain method of the search for the deuteron electric dipole moment in a storage ring with imperfections. The method differs from the one~\cite{Farley,AGS_proposal} that was previously proposed in that we use the measurement of the frequency of the total EDM and MDM signal, as opposed to the value of the vertical component of spin. This method allows systematic errors to be reduced to a level at which the lower limit of detection of the assumed EDM can be as low as $\sim 10^{-29}\div 10^{-30}e\cdot cm$. 

The authors would like to thank the JEDI collaboration for their scientific discussions and for their support in helping to make this work a success.

\end{document}